\documentclass[aps,prb,twocolumn,superscriptaddress,showpacs]{revtex4-1}
\usepackage{graphicx}
\usepackage{textcomp}
\usepackage{amsmath}
\usepackage{color}

\begin{document}

\title{Analytical description of ballistic spin currents and torques in magnetic tunnel junctions}

%\author{M.~Chshiev$^{1,2}$, A.~Manchon$^3$, A.~Kalitsov$^1$, N.~Ryzhanova$^4$, A.~Vedyayev$^4$, W.~H.~Butler$^2$, B.~Dieny$^1$}

\author{M.~Chshiev}
\affiliation{Univ. Grenoble Alpes, INAC-SPINTEC, F-38000 Grenoble, France; CNRS, SPINTEC, F-38000 Grenoble, France; and CEA, INAC-SPINTEC, F-38000 Grenoble, France}
\affiliation{MINT Center, University of Alabama, Tuscaloosa, AL, 35487-0209, USA}
\author{A.~Manchon}
\affiliation{Physical Science and Engineering Division, King Abdullah University of Science and Technology (KAUST), Thuwal 23955-6900, Saudi Arabia}
\author{A.~Kalitsov}
\affiliation{Univ. Grenoble Alpes, INAC-SPINTEC, F-38000 Grenoble, France; CNRS, SPINTEC, F-38000 Grenoble, France; and CEA, INAC-SPINTEC, F-38000 Grenoble, France}
\affiliation{MINT Center, University of Alabama, Tuscaloosa, AL, 35487-0209, USA}
\author{N.~Ryzhanova}
\affiliation{Department of Physics, Moscow Lomonosov State University, Moscow 119991, Russia}
\author{A.~Vedyayev}
\affiliation{Department of Physics, Moscow Lomonosov State University, Moscow 119991, Russia}
\author{N.~Strelkov}
\affiliation{Department of Physics, Moscow Lomonosov State University, Moscow 119991, Russia}
\author{W.~H.~Butler}
\affiliation{MINT Center, University of Alabama, Tuscaloosa, AL, 35487-0209, USA}
\author{B.~Dieny}
\affiliation{Univ. Grenoble Alpes, INAC-SPINTEC, F-38000 Grenoble, France; CNRS, SPINTEC, F-38000 Grenoble, France; and CEA, INAC-SPINTEC, F-38000 Grenoble, France}

\date{\today}

\begin{abstract}
In this work we demonstrate explicit analytical expressions for both charge and spin currents which constitute the 2x2 spinor in magnetic tunnel junctions with non-collinear magnetizations under applied voltage. The calculations have been performed within the free electron model in the framework of the Keldysh formalism and WKB approximation. We demonstrate that spin/charge currents and spin transfer torques are all explicitly expressed through only three irreducible quantities, without further approximations. The conditions and mechanisms of deviation from the conventional sine angular dependence of both spin currents and torques are evidenced and discussed. It is shown in the thick barrier approximation that all tunneling transport quantities can be expressed in an extremely simplified form via Slonczewski spin polarizations and the newly introduced effective spin averaged interfacial transmission probabilities and effective out-of-plane polarizations at both interfaces. It is proven that the latter plays a key role in the emergence of perpendicular spin torque as well as for the angular dependence character of all spin and charge transport considered. It is demonstrated directly also that for any applied voltage, the parallel component of spin current at the FM/I interface is expressed via collinear longitudinal spin current components. Finally, STT behavior is analyzed in a view of transverse characteristic length scales for spin transport.
\end{abstract}

%\keywords{}
\maketitle

%\section{Introduction}

Interest in spintronics has been strongly accentuated by the discovery of current-induced magnetization switching (CIMS) caused by spin transfer torque (STT) in both metallic multilayers and tunnel junctions~\cite{Slonc96,Berger,huai1,huai2,Fuchs,Ohno,chapter1,chapter2}. In the ballistic transport regime, such switching is caused by STT resulting from the non-conservation of transverse components of spin currents~\cite{Slonc96}. On the other hand, in magnetic tunnel junctions (MTJs), the spin-dependent charge currents determine the tunneling magnetoresistive (TMR) properties. Consequently, the key for understanding the fundamental mechanisms underlying both STT and TMR in MTJs with non-collinear magnetizations is the understanding of the fundamental quantum properties of both spin and charge currents~\cite{chapter2}.

Since the pioneering work of Slonczewski, a number of theoretical studies have addressed the microscopic details of STT in MTJs, using various approaches for calculating spin and charge transport.  These include  the transfer matrix formalism~\cite{slonc05,slonc07}, the tight-binding approach~\cite{kalitsov06,theo,Chshiev,kalitsov09,kioussis}, the free electron approach~\cite{slonc89,wil,xiao,manchon}, and approaches based on first principles calculations of the electronic structure~\cite{heiliger,guo}. It is now well established that for elastic tunneling in MTJs, STT possesses two components of the form:
\begin{eqnarray}\label{eq:stt}
{\bf T}_{||}&=&(a_1V+a_2V^2){\bf M}_R\times({\bf M}_R\times {\bf M}_L),\\\label{eq:op}
{\bf T}_{\bot}&=&(b_0+b_1V+b_2V^2){\bf M}_R\times {\bf M}_L,
\end{eqnarray}
where ${\bf M}_L$ and ${\bf M}_R$ are the magnetization directions of the pinned and free layers, respectively. Several properties can be outlined for the STT components. For instance, it has been shown that $a_2$ in Eq.~(\ref{eq:stt}) vanishes for the case of half-metallic electrodes~\cite{Chshiev,kalitsov09} while $b_1$ tends to zero in Eq.~(\ref{eq:op}) for symmetric MTJs yielding quadratic bias voltage behavior~\cite{theo,Chshiev,manchon,wil,xiao} as confirmed later experimentally~\cite{sankey,kubota}. The roles of inelastic scattering, structural asymmetries and material compositions have also been theoretically investigated~\cite{swstt,magnon,manchon2,manchon3,Kalitsov2013,pauyac,merodio,useinov} resulting in the forms displayed in Eqs. (\ref{eq:stt})-(\ref{eq:op}) with some of them supported by a number of recent experiments~\cite{deac,petit,li,sun,oh,chanthbouala}. In certain cases the perpendicular torque ${\bf T}_{\bot}$ may oscillate with the bias voltage~\cite{kioussis} indicating that Eq. ({\ref{eq:op}) is only a low bias approximation.

\begin{figure}[htb]
\begin{center}
\includegraphics[width=0.7\columnwidth]{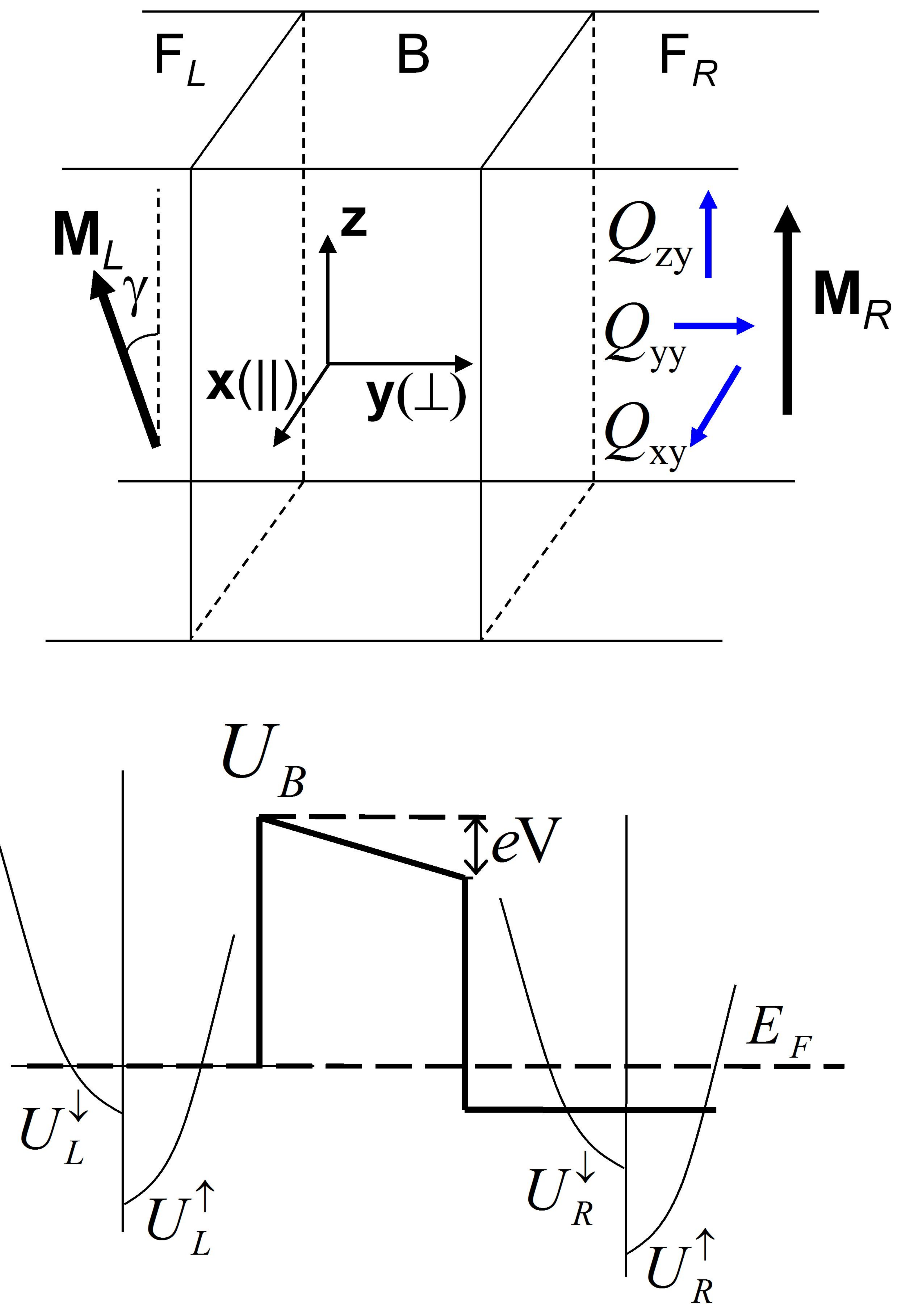}% Here is how to import EPS art
\caption{\label{fig1} (top) Schematic structure of the MTJ, consisting of left and right semi-infinite FM leads separated
by a thin nonmagnetic insulating barrier. The magnetization ${\bf M}_R$ of the right FM lead is along $z$, whereas the
magnetization ${\bf M}_L$ of the left lead is rotated by an angle $\gamma$ around the y axis with respect to ${\bf M}_R$. 
(bottom) Schematic illustration of the potential profile, where $U^{\uparrow}_{L(R)}$, $U^{\downarrow}_{L(R)}$, and $U_B$ are the potentials of the majority and minority bands in the left(right) FM leads, and the barrier, respectively. The lower dashed line indicates the Fermi level in equilibrium.}
\end{center}
\end{figure}
        
Due to the cumbersome form of the actual expressions, most of the models proposed up to now rely on numerical simulations~\cite{theo,Chshiev,kioussis,wil,manchon}. Therefore, no transparent formulas are available to qualitatively describe the torques in MTJs. In this work we demonstrate explicit analytical expressions for both charge and spin currents which constitute the $2\times2$ current matrix in magnetic tunnel junctions with non-collinear magnetizations of two ferromagnetic electrodes, F$_L$ and F$_R$ with angle $\gamma$ between them. The electrodes are  separated by an insulator (B) and there is an applied voltage V. As shown in Fig.~\ref{fig1}, the interface is taken to be perpendicular to the $y$-direction and the magnetization of the free (F$_R$) layer is assumed to be in the $z$-direction . 

The expressions derived here can form a good basis for understanding the physics of TMR, interlayer exchange coupling and STT in the case of both symmetric and asymmetric MTJs. The calculations have been performed within the free electron model using the non-equilibrium Green function technique in the framework of the Keldysh formalism and WKB approximation. They extend the previous approximate expressions reported in the literature~\cite{chapter1,slonc89,manchon,manchon2}. The expressions presented here, however, have a very compact form since they are expressed through only three irreducible quantities without further approximations. This allows easy implementation in commercial software by applying straightforward integration rules~\cite{Supplementary}. Moreover, in the limit of thick barriers all non-collinear transport quantities are expressed in an extremely simplified form via Slonczewski spin polarizations and the newly introduced {\it effective spin averaged interfacial transmission probabilities} and {\it effective out-of-plane polarizations} at both interfaces. We demonstrate that the latter reflects the degree of spin mistracking which gives rise to the perpendicular STT term and determines the angular dependence  of STT and TMR.

Prior to entering into the details of the obtained results, it is important to point out the main limitations of the formulas reported in this letter. It is well established that the complex band structure of MgO-based MTJs has an important impact on microscopic transport properties, such as STT and TMR, and can be accounted for through first-principle calculations~\cite{heiliger}. This technique allows for considering a realistic density of states, as well as the symmetry characters of the tunneling electrons, resulting in important properties such as resonant interfacial states and so on. Although this technique, combined with the Non-Equilibrium Green's Function (NEGF) formalism, provides a realistic bias dependence of  STT~\cite{heiliger}, a simpler model using tight-binding theory has proven to be sufficient for predicting and describing the essential characteristics of the spin transfer torques in MTJs, (including barriers other than   MgO~\cite{theo,Chshiev,kalitsov09,kioussis}). In this case, the interfacial density of states is modeled by a closed-form band dispersion relation which allows for varying the effective band filling, giving rise to unexpected bias dependencies of the torque components~\cite{theo,Chshiev,kalitsov09,kioussis,manchon}. 

In the present article, we choose the free electron model in which the dispersion is parabolic.  However, for low band-filling, tight-binding bands are well approximated by free-electron dispersion.  In particular, for the important case of bcc (Co)Fe, the dispersion relation of the $\Delta_1$ band is similar to free electron dispersion. This approach provides an efficient, compact and transparent qualitative description of ballistic tunneling which may be useful and important  in view of potential new technological applications of spin transfer torque.  Finally, this approach uses the standard quantum mechanical procedure of matching electron wave functions at the interfaces of the MTJ, a procedure that is formally equivalent to the NEGF method developed in Refs.~\citenum{kalitsov06,theo,Chshiev,kalitsov09}.

%This paper is organized as follows: Section~\ref{methodology} describes the methodology and derivations...

%\section{Methodology and derivation}\label{methodology}

Spin and charge transport  across a non-collinear MTJ are represented by elements of a 2x2 current matrix in spin space which can be written as
\begin{eqnarray}\label{2x2}
\hat J=\left(
\begin{array} {ccc} J^{\uparrow \uparrow} & J^{\uparrow \downarrow} \\
J^{\downarrow \uparrow} & J^{\downarrow \downarrow} \end{array} \right ) =
\left(
\begin{array} {ccc} \Lambda^{\uparrow(\uparrow)}+\Lambda^{\uparrow(\downarrow)} & \Xi^{\uparrow(\uparrow)}+\Xi^{\uparrow(\downarrow)} \\
\Xi^{\downarrow(\uparrow)}+\Xi^{\downarrow(\downarrow)} & \Lambda^{\downarrow(\uparrow)}+\Lambda^{\downarrow(\downarrow)}\end{array} \right )
\end{eqnarray}
where $\Lambda$ and $\Xi$ are described in detail in the Appendix. The matrix above defines the required charge current and spin current tensor components $Q_{ij}$ (with indices $i$ and $j$ being in spin and real space, respectively)\textbf{\textcolor{red}{}} using the identity, $\hat I,$ and the Pauli matrices, ($\hat\sigma_x$, $\hat\sigma_y$, $\hat\sigma_z$).  The diagonal elements of (\ref{2x2}) can be used to express the total charge and longitudinal spin currents as, $J_e=-(|e|/\hbar)\mathrm{Tr}(\hat J\, \hat I)$ and $Q_{zy}=\mathrm{Tr}(\hat J\hat \sigma_z)/2$, respectively. The non-diagonal elements $J^{\uparrow \downarrow (\downarrow \uparrow)}=Q_{xy}\pm\mathrm{i}Q_{yy}$ comprise transverse spin current tensor components which are extracted using $\hat\sigma_x$ and $\hat\sigma_y$, i.e. $Q_{xy}=\mathrm{Tr}(\hat J\hat \sigma_x)/2$ and $Q_{yy}=\mathrm{Tr}(\hat J\hat \sigma_y)/2$. In the following, the second (real space) index in the spin current expressions will be omitted.  Only the index  pertaining to spin space will be retained.

The spin-dependent wave vectors in the $i$-th electrode are denoted by $k^{\sigma}_i$ ($i$ is ``L'' or ``R'', $\sigma$ is ``$\uparrow$''(+) or ``$\downarrow$''(-)) and the wave vector inside the barrier is denoted by  $q(y).$ Detailed expressions for these quantities are given in the Appendix [Eqs.~(\ref{eq:k}) and (\ref{eq:q})].
In order to give an explicit account of the torques and current, we define three irreducible factors, $P_i=(k_i^\uparrow-k_i^\downarrow)/(k_i^\uparrow+k_i^\downarrow)$, $\alpha_i=(q_i^2-k_i^\uparrow k_i^\downarrow)/(q_i^2+k_i^\uparrow k_i^\downarrow)$ and $\eta_i=q_i(k_i^\uparrow+k_i^\downarrow)/(q_i^2+k_i^\uparrow k_i^\downarrow).$ Note that $q_i$ should be replaced by $q_i/m_{\mathrm{eff}}$ where $m_{\mathrm{eff}}=m^*/m_e$ when $m_{\mathrm{eff}}\neq 1$. The first two factors are referred to as Stearns' polarization \cite{stearns} and Slonczewski's factor~\cite{slonc89} with their product giving Slonczewski's spin polarization $P^S_i=P_i\alpha_i$ for both interfaces. For reasons clarified further later,  $P^S_i$ can be viewed as an effective in-plane polarization while the product of $P_i$ and $\eta_i$,  $P^\eta_i=P_i\eta_i,$ will be referred as an effective out-of-plane polarization.

We will now proceed to the expressions for spin transfer torques as well as spin and charge currents which, even in the general case, can be  conveniently expressed using only $P_i$, $\alpha_i$ and $\eta_i$ [see Eqs.~(\ref{eq:tip})-(\ref{Den}) and (\ref{Qx})-(\ref{chcurr})]. After that, we will show that for a barrier that is sufficiently thick and high, these  expressions take extremely simple and clear forms which can be expressed straightforwardly using only $P^S_i$, $P^\eta_i$ and the effective spin averaged interfacial transmission probabilities defined as $T_i=\eta_i/(\eta_i^2+\alpha_i^2)$ [Eqs.~(\ref{eqs:tip})-(\ref{eqs:long})]. Note that the latter represents the effective transmission probability through interfaces for both spin channels and in this sense is different from the one traditionally used which is expressed for each spin channel separately~\cite{slonc05,ZhangButler,BelaschTsymbal}.

In the absence of spin relaxation and spin-orbit coupling, the spin transfer torque, ${\bf T}$, can be written as,  ${\bf T}=-\nabla {\bf Q}$, where the real-space part of the spin current tensor (Fig.~\ref{fig1}) is contracted by the divergence operator. Taking into account the vanishing of the transverse spin current 
far from the interface~\cite{kalitsov06,theo,Chshiev}, the in-plane and perpendicular torques exerted by the left layer on the right layer can be expressed through the interfacial spin current integrand in the barrier $T_{||(\bot)}=Q_{x(y)}^B$ and are given by the following expressions:
\begin{equation}\label{eq:tip}
T_\parallel=\frac{4\sin\gamma}{|Den|^2} P_L\left(2\alpha_R-\alpha_L[E_n^2+E_n^{-2}] \right)[f_L - f_R]
\end{equation}
\begin{equation}\label{eq:top}
T_\perp=-\frac{4\sin\gamma}{|Den|^2} P_LP_R\left(\alpha_L\eta_Rf_L +\alpha_R\eta_Lf_R  \right)[E_n^2-E_n^{-2}]
\end{equation}
\begin{align}\label{Den}
\begin{split}
|Den|^2=\frac{1}{\eta_L\eta_R}&\Biggl\{ \left( \alpha_L\eta_R+\alpha_R \eta_L\right)^2 \left( E_n^2-E_n^{-2}\right)^2 \Biggr.\\
&+\Bigl[\left( \alpha_L\alpha_R-\eta_L\eta_R \right)(E_n^2+E_n^{-2}) \Bigr.\\
&\Biggl.\Bigl.- 2\alpha_L\alpha_R+2P_LP_R\eta_L\eta_R\cos{\gamma}\Bigr]^2\Biggr\}
\end{split}
\end{align}
where $E_n$ is defined in Eq.~(\ref{eq:En}) and $f_L$, $f_R$ are the standard Fermi distribution functions shifted by $\pm eV/2$, respectively . Note that all transport quantities presented here are functions of $E$ and $\kappa^2=k_x^2+k_y^2$, so in order to obtain numerical results, the above explicit formulas need to be integrated over the Fermi sphere, considering $2m E/\hbar^2+\kappa^2\in[-\infty,E_F]$. A detailed calculation procedure is given in Appendix. 
\begin{figure}[t]
	\includegraphics[width=1\columnwidth]{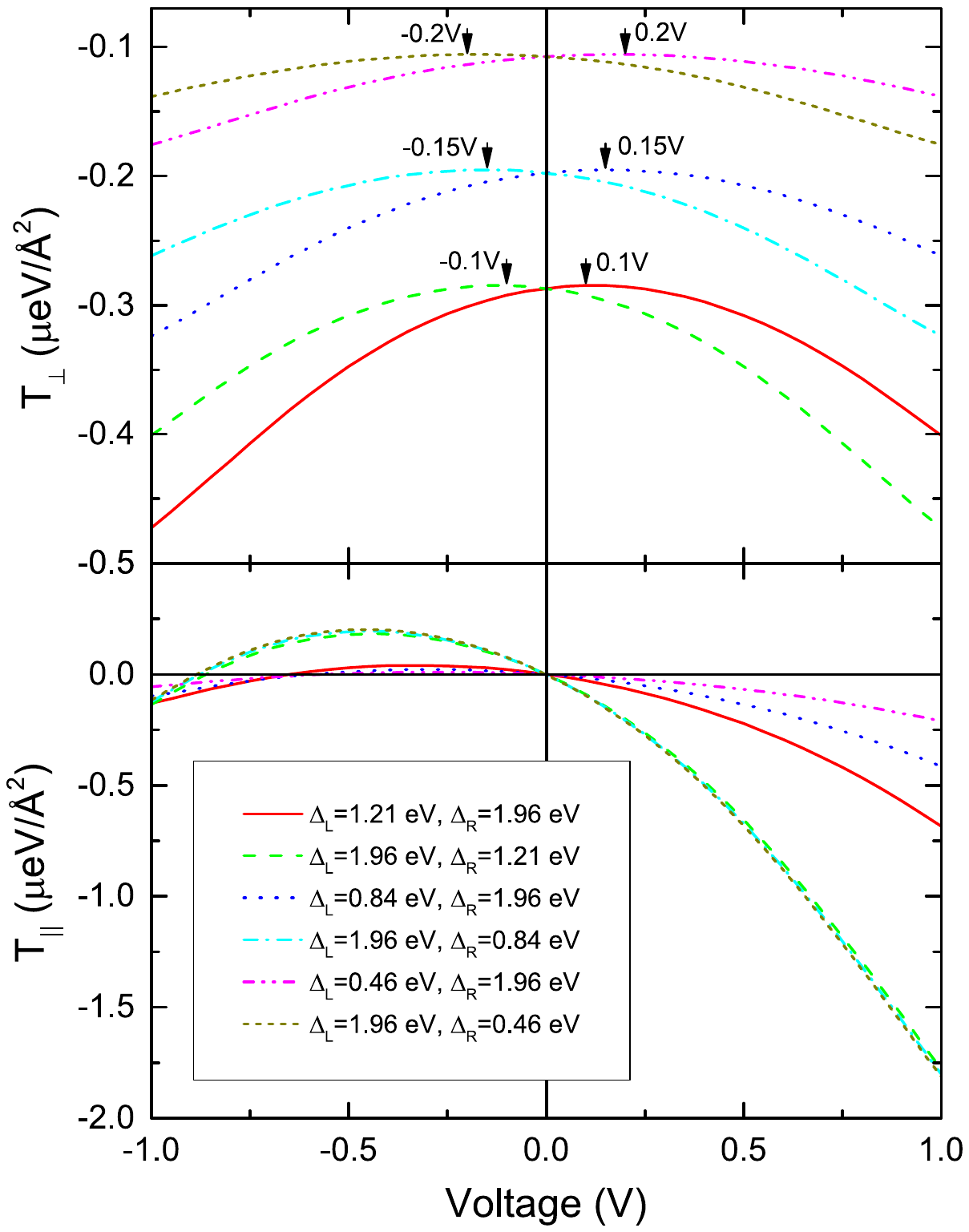}
	\caption{\label{fig:Tper_asym} The field-like $T_\perp$ and parallel $T_\parallel$ torques as a function of applied bias in an asymmetric MTJ. $U_B=1$~eV, $\Delta_0=2.62$eV, $m_{\mathrm{eff}}=0.4$, $\gamma=\pi/2$, $d=7$~\AA.\label{fig:displacement}}
\end{figure}
\begin{figure*}[t]
	\includegraphics[width=2\columnwidth]{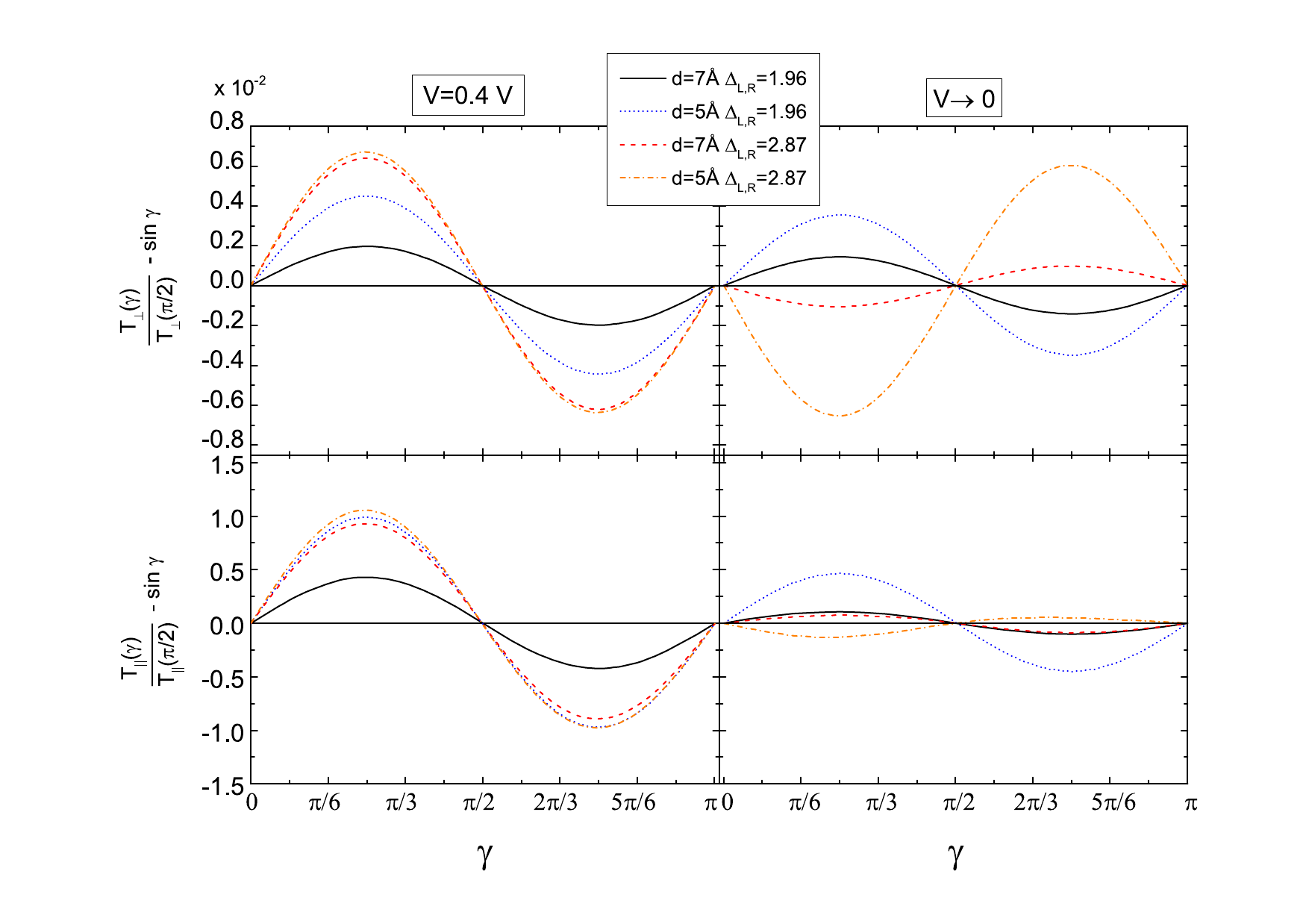}
	\caption{Angular dependence of an universal deviation function given by Eq.~(\ref{devfunc}) for $T_\perp$ (top panels) and $T_\parallel$ (bottom panels) for different values of thickness $d$, exchange splittings  $\Delta_{L,R}$ and applied voltages V. The parameters are: $\Delta_0=2.62$eV, , $U_B=0.5~$eV, $m_{\mathrm{eff}}=0.4$. \label{fig:deviation}}
\end{figure*}

It is worth commenting on the above equations. First, note that the perpendicular torque, $T_\bot$ (Eq.~\ref{eq:top}) is not modified under $R\leftrightarrow L$ exchange. Therefore, for the case in which the left and right electrodes are equivalent, the term $\alpha_L\eta_Rf_L +\alpha_R\eta_Lf_R$ becomes {\em symmetric} in the bias voltage. It can be shown straightforwardly by series expansion in voltage that this situation yields a perpendicular torque on the form~\cite{oh} $\sum_n b_{2n}V^{2n}$ which agrees with to the symmetric (even parity) bias dependence obtained numerically~\cite{kioussis}. When structural asymmetries are present in the junction, so that the left and right interfaces are no longer equivalent, the perpendicular torque displays linear and higher order antisymmetric components~\cite{manchon2,oh} taking the form $\sum_n b_{n}V^{n}$ as shown in Fig.~\ref{fig:displacement} and observed experimentally~\cite{oh}. Interestingly, the curve peak displacement is proportional to the difference between the exchange splitting of the right and left FM layers. Note also, that  the $n=0$ term in these expansions for $T_\bot$ as well as in Fig.~\ref{fig:Tper_asym} represents the inter-layer exchange coupling~\cite{slonc89,IEC_exp,IEC_the,Yang}. On the contrary, the in-plane torque, $T_{||}$, given by Eq. (\ref{eq:tip}), does not have this type of structural symmetry and therefore displays a wide range of behavior as a function of applied bias even in a symmetric junction~\cite{theo,Chshiev,kalitsov09,manchon}. It is interesting to note in Fig.~\ref{fig:Tper_asym} that the in-plane torque is almost  insensitive to the exchange splitting of the right layer and depends instead on that of the polarizer in perfect agreement with Eq.~(\ref{eq:tip}) where it is defined indeed by the Stearns and more importantly by Slonczewski in-plane polarizations of the left layer.

The angular dependence on the torques displayed in Eqs.~(\ref{eq:tip}-\ref{Den}) is one of the important results of this work. It is apparent that the deviation from the conventional $\sin\gamma$ dependence is contained in the denominator $|Den|^2$, where the angular dependence is given by the term proportional to $\eta_L\eta_RP_LP_R\cos\gamma$. For thicker and/or higher barriers this term is negligible compared to $E_n^4$ so that the denominator is independent on the angle resulting in the torque being simply proportional to  $\sin\gamma$ in this limit. However, when the barrier is made thinner or lower, the denominator terms with $\cos\gamma$ can no longer be neglected and the angular dependence deviates from the standard $\sin\gamma$ form. One can introduce a universal deviation function for both torques of the following form:
\begin{widetext}
        \begin{eqnarray}\label{devfunc}
        \frac{{T\left( \gamma  \right)}}{{T\left( {\pi /2} \right)}} - \sin \gamma  =  - \frac{{4\sin \gamma P_L^\eta P_R^\eta }}{{{{\left| {Den} \right|}^2}}} \times \left\{ {\left[ {{\alpha _L}{\alpha _R}{{\left( {{E_n} - E_n^{ - 1}} \right)}^2} - {\eta _L}{\eta _R}\left( {E_n^2 + E_n^{ - 2}} \right)} \right]\cos \gamma  + {P_L^\eta}{P_R^\eta}{{\cos }^2}\gamma } \right\}
        \end{eqnarray}
\end{widetext}
A major role in these deviations is played by the effective out-of-plane polarization $P_i^\eta$ through the $P^\eta_LP^\eta_R\cos\gamma$ term.
  Fig.~\ref{fig:deviation} represents the deviation function behavior for both torques for different barrier thicknesses, exchange splittings and applied voltages. One can see that the angular dependence of the deviation function for all cases shows a $\sin2\gamma$ form governed by the first term in Eq.~(\ref{devfunc}). The magnitude of the deviations is of the order of 1\% or less for barrier thickness and height as low as 5~\AA~and 0.5~eV, respectively, indicating that the usual $\sin\gamma$-dependence of STT is quite robust for a wide range of materials including MgO or AlOx barriers, consistent with the thick barrier approximation which will be introduced below. 

The parallel and perpendicular spin currents in the right FM electrode are given by the following expressions:
\begin{widetext}
        \begin{eqnarray}\label{Qx}
        Q_{x}^{R} = \frac{4\sin\gamma}{|Den|^2} P_L&\left[ P_R\eta_R\alpha_L[E_n^2-E_n^{-2}] \sin\{\Delta k_Ry\}\right. 
        +\left(2\alpha_R-\alpha_L[E_n^2+E_n^{-2}] \right) \left. \cos\{\Delta k_Ry\}\right][f_L - f_R]
        \end{eqnarray}
        \begin{eqnarray}\label{Qy}
        \begin{split}
        Q_{y}^{R} = \frac{4\sin\gamma}{|Den|^2} P_L&\Bigg\{\bigg[\left(2\alpha_R-\alpha_L[E_n^2+E_n^{-2}] \right)\sin\{\Delta k_R y\}
        -P_R\eta_R\alpha_L[E_n^2 - E_n^{-2}] \cos\{\Delta k_R y\}\bigg][f_L - f_R] \\
        &+P_Rf_R \bigg[ \left(\left(\eta_L\eta_R-\alpha_R\alpha_L\right)[E_n^2+E_n^{-2}]+ 2(1-\eta_L\eta_RP_LP_R\cos\gamma)\right)\sin\{\Sigma k_Ry\}\bigg.\bigg.\\
        &-\left(\eta_L\alpha_R+\eta_R\alpha_L\right)[E_n^2-E_n^{-2}]\cos\{\Sigma k_Ry\} \bigg ] \Bigg\}
        \end{split}
        \end{eqnarray}
\end{widetext}
where $\Delta k_i=k_i^\uparrow - k_i^\downarrow$ and $\Sigma k_i=k_i^\uparrow+k_i^\downarrow$. The beating described in Ref. \citenum{manchon} is now displayed explicitly in the $\cos\{(k_R^\uparrow + k_R^\downarrow)y\}$, $\cos\{(k_R^\uparrow - k_R^\downarrow)y\}$, $\sin\{(k_R^\uparrow + k_R^\downarrow)y\}$, and $\sin\{(k_R^\uparrow - k_R^\downarrow)y\}$ terms. The corresponding local spin transfer torques in the right FM electrode can be obtained by taking the derivative of Eqs.~(\ref{Qx}) and (\ref{Qy}) with respect to the $y$-coordinate. 
As one can see in Fig.~\ref{fig:T_y_sym} both STT terms within the right FM electrode oscillate and decay as a function of distance from the B$|$F$_R$ interface in agreement with previous reports for MTJs~\cite{kalitsov06,manchon} and metallic spin valves~\cite{Stiles,Slon2002,Waintal2010}. It is interesting to note that the period of oscillations $\lambda_L$, (which is related to the  Larmor spin precession length, $l_L$, by a factor $2\pi$), is different under positive and negative applied voltages [cf. Figs.~\ref{fig:T_y_sym}(a) and (b)]. This is due to the asymmetric voltage dependence of $\Delta k_R$ which defines the oscillation wavelength $\lambda_L$ as 2$\pi/\Delta k_R$. Indeed, as shown by the solid lines in the inset of Fig.~\ref{fig:T_y_sym}, $\Delta k_R$ is larger (smaller) for negative (positive) applied voltage resulting in smaller (larger) $\lambda_L$ in Fig.~\ref{fig:T_y_sym}(a) and (b), respectively. The oscillations decay with an exponential envelope function $~e^{-(y-d)/\lambda_d}$ which results from the dephasing due to integration over the in-plane momentum $\kappa$ within the tunneling cone~\cite{manchon}. Here, $\lambda_d$, indicates the characteristic transverse spin decay length~\cite{note}. The latter is strongly dependent on the applied voltage, varying from  $\sim$20~\AA\ to $\sim$7.5~\AA\ as the bias changes from  +1~V  to -1~V  as shown in Fig~\ref{fig:T_y_sym}(a) and (b), respectively. Similar behavior is observed  when the right FM electrode becomes a half-metal [Fig.~\ref{fig:T_y_half}] with both $\lambda_L$ and $\lambda_d$ being smaller compared to the aforementioned case of non-half-metallic FM electrode [cf.~Figs.~\ref{fig:T_y_half} and~\ref{fig:T_y_sym}]. Such behavior can again be explained  by a change in the dependence of $\Delta k_R$ as a function of applied voltage when one compares insets in Figs.~\ref{fig:T_y_half} and~\ref{fig:T_y_sym}. In particular, the presence of the imaginary part of $\Delta k_R$ in all ranges of negative voltages yields a significantly stronger decay of oscillations as displayed in Fig.~\ref{fig:T_y_half}(b).

\begin{figure}[t]
        \includegraphics[width=\columnwidth]{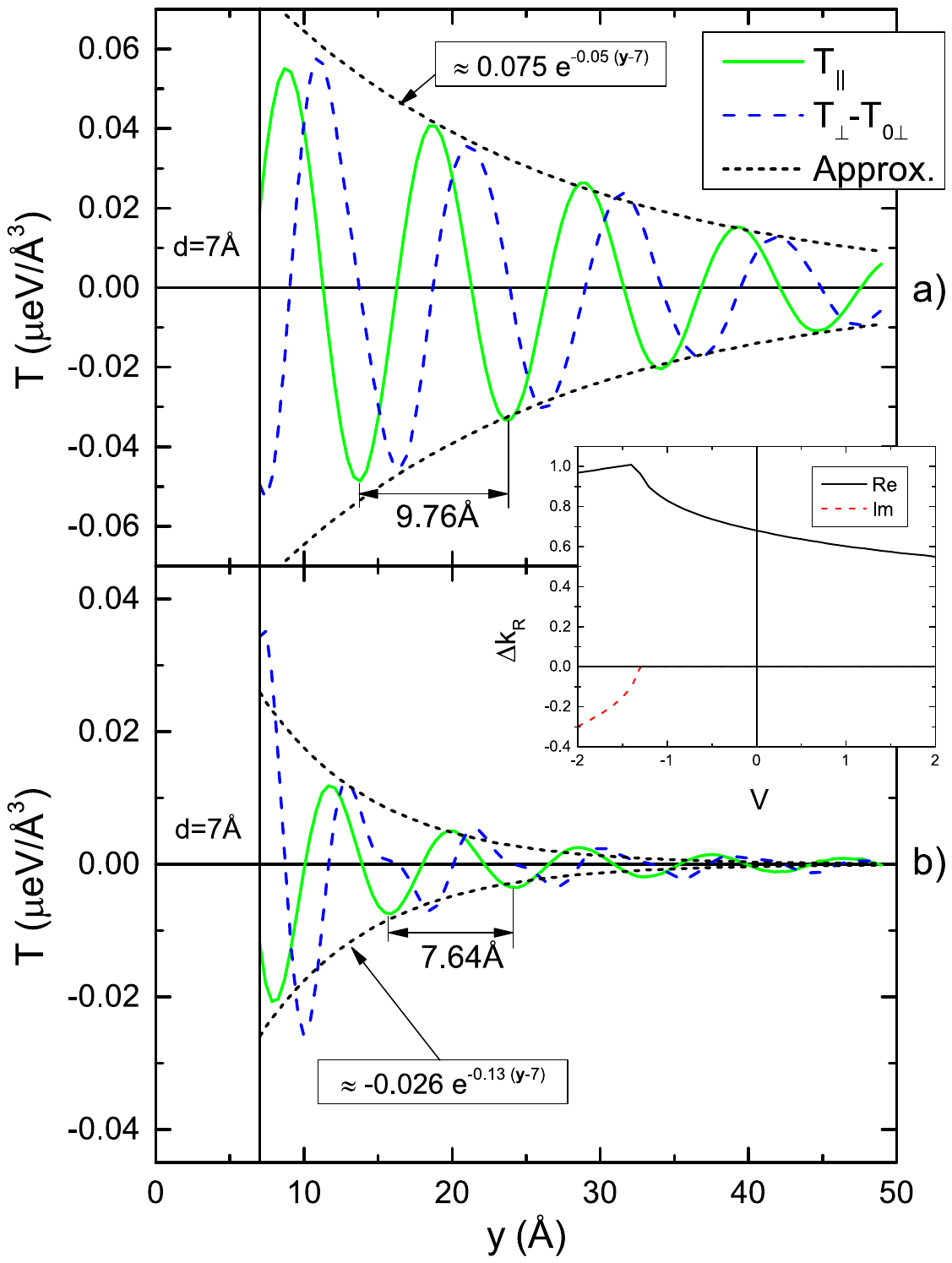}
        \caption{Distribution of the voltage induced in-plane and out-of-plane STT terms in the right FM electrode of a symmetric MTJ for (a) V=+1~V and (b) V=-1~V applied voltage. The parameters are: $\Delta_0=2.62$eV, $\Delta_L=\Delta_R=1.96$eV, $U_B=3$eV, $m_{\mathrm{eff}}=0.4$, $\gamma=\pi/2$, $d=7$\AA. $T_{0\bot}$ indicates the field-like torque at zero voltage. The inset shows a voltage dependence of the real and imaginary parts of $\Delta k_R=k^\uparrow_R-k^\downarrow_R$. \label{fig:T_y_sym}}
\end{figure}
\begin{figure}[t]
        \includegraphics[width=\columnwidth]{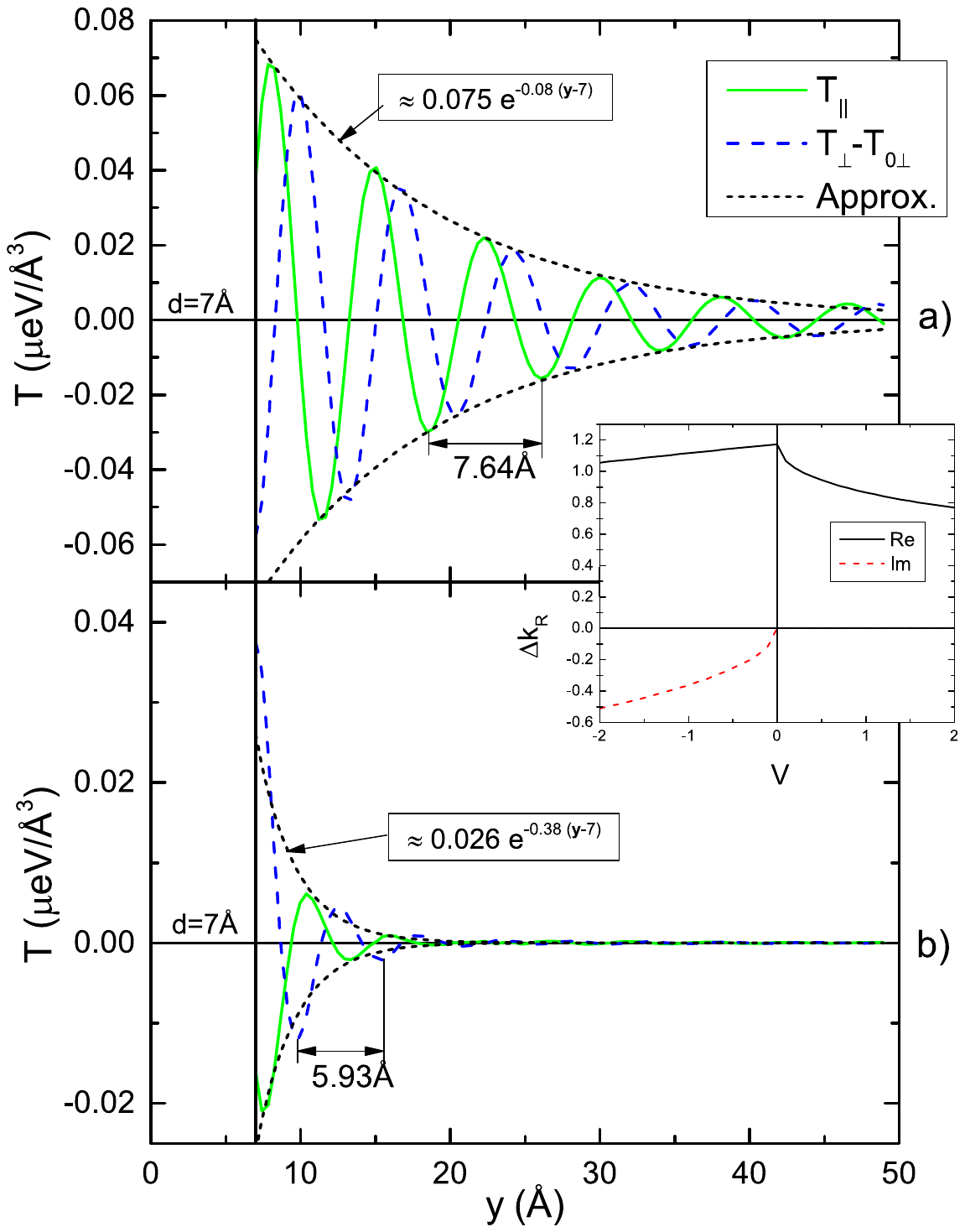}
        \caption{The same as in Fig.~\ref{fig:T_y_sym} for case in which the right FM electrode is a half-metal. Exchange splittings are $\Delta_0=\Delta_R=2.62$eV and $\Delta_L=1.96$eV while other parameters unchanged.~\label{fig:T_y_half}}
\end{figure}

Next, we give exact expressions for the longitudinal spin current $Q_{z}$ and charge current $J_e$ which are conserved through the barrier and the right electrode and can respectively be expressed as:
\begin{widetext}
        \begin{align}\label{chcurr}
        \begin{split}
        J_e = -\frac{8}{|Den|^2}&\bigg\{\left(1+\alpha_L\alpha_RP_LP_R\cos\gamma \right)[E_n^2 + E_n^{-2}]\\
        &-2\left(\alpha_L\alpha_R+P_LP_R\cos\gamma-\right.
        \left.\eta_L\eta_R(1-\alpha_LP_L)(1-\alpha_RP_R)\right) \bigg\}[f_L-f_R] \\
        Q_{z}^B=-\frac{4}{|Den|^2}&\bigg\{\left(\alpha_RP_R+\alpha_LP_L\cos\gamma \right)[E_n^2 + E_n^{-2}]
        +2\left(\alpha_LP_R+\alpha_RP_L\cos\gamma \right)\bigg\}[f_L-f_R]
        \end{split}
        \end{align}
\end{widetext}

One of the most important results of this work is the form of the aforementioned expressions when the barrier is thick, i.e. $E_n^2\gg 1$, and as justified above by small deviations from sine angular dependencies of STT given by Eq.~(\ref{devfunc}) and shown in Fig.~\ref{fig:deviation}. In this case, the formulas can be written using only $P^S_i$, $P^\eta_i$ and $T_i$ and take a very simple form:
\begin{equation}\label{eqs:tip}
T_{||}=-4 T_L T_R P_L^S E_n^{-2}[f_L - f_R]\sin\gamma
\end{equation}
\begin{equation}\label{eqs:top}
T_{\bot}=-4 T_L T_R \left(P_L^S P_R^\eta f_L + P_R^S P_L^\eta f_R\right) E_n^{-2}\sin\gamma
\end{equation}
\begin{equation}\label{eqs:cur}
J_e=-8 T_L T_R \left(1+P_L^S P_R^S\cos\gamma\right) [f_L - f_R] E_n^{-2}
\end{equation}
\begin{equation}\label{eqs:long}
Q_{z}=-4 T_L T_R \left(P_R^S + P_L^S\cos\gamma\right)[f_L - f_R] E_n^{-2}
\end{equation}
One can note that the previously introduced in-plane (Slonczewski's) polarization, $P_i^S$, and the out-of-plane polarization, $P_i^eta$, play a very different role. While the former defines the magnitude of the TMR and both components of the spin torque [see Eqs.~(\ref{eqs:tip})-(\ref{eqs:long})], the latter participates only to the out-of-plane torque, $T_{\bot}$ [see Eq.~(\ref{eqs:top})]. This is the second important role of the out-of-plane polarization, besides being responsible for the aforementioned angular deviation. 
%Let us now discuss and clarify physical mechanisms of such a decisive role of $P^\eta_i$ for $T_{\bot}$ and angular dependences of spin and charge transport. 
In fact, $P^\eta_i$ is decisive for $T_{\bot}$ since it accounts for the degree of out-of-plane precession  at the interfaces for a spin initially polarized along ${\bf M}_L$ which ensures the appearance of $Q_y$ giving rise to the perpendicular torque (see Fig.~\ref{fig1}). Interestingly, it can be shown in the approximation where $q_i^2 \gg k_i^{\uparrow}k_i^{\downarrow}$, that $P^\eta_i$ and $P^S_i$ can be literally assigned respectively to the sine and cosine of the out-of-plane precession angle $\phi$ at the interfaces which in this case is very small. This is the reason that we named $P^\eta_i$ and $P^S_i$ the in-plane and out-of-plane interfacial polarizations, respectively. The situation changes as the  electron energy becomes closer to the barrier height, i.e. when $q_i$ no longer dominates the geometrical mean of $k_i^{\uparrow}$ and $k_i^{\downarrow}$. In this case, the electron spin starts to ''precess'' or get reoriented prior to arrival at the right interface since it begins to have enough energy to interfere with its reflected part leading to a much stronger decrease of $P^S_i$ compared to $P^\eta_i$ affecting thereby the  TMR and $T_{\|}$ amplitudes. Finally, when the barrier becomes low and thin, the terms in the denominator given by Eq.~(\ref{Den}) accounting for multiple ``interferences'' of transmitted and reflected non-collinear evanescent states due to their strong overlap between two interfaces, result in further modulation of the out-of-plane spin component taken into account through $P^\eta_i$. This leads to the aforementioned deviations from the standard sine angular dependences of spin transfer torques and the standard cosine angular dependence of the   charge and spin currents. 

Another important result is that it is straightforward to show using Eqs.~(\ref{eqs:tip}) and (\ref{eqs:long}) that the parallel spin current in the barrier region which represents the total in-plane spin torque deposited in the right FM electrode can be expressed using the longitudinal spin current for parallel and antiparallel components respectively as~\cite{theo}
\begin{equation}\label{collformula}
T_{\|}=Q_{x}=\frac{Q_{z}(0)- Q_{z}(\pi)}{2} {\bf M}_{R} \times ({\bf M}_{R}\times {\bf M}_{L}).
\end{equation}
Finally, one can get a corrected generalized expression for spin torque efficiency by dividing Eqs.~(\ref{eq:tip}) and (\ref{eq:top}) by Eq.~(\ref{chcurr}) without forgetting to subtract the zero voltage part which accounts for the equilibrium exchange coupling through $T_\bot$. For instance, in the limit of a thick/high barrier, using Eqs.~(\ref{eqs:tip}) and (\ref{eqs:cur}) the in-plane torque efficiency becomes  $T_\|/J_e=-(1/2)P^S_L\sin\gamma/(1+P^S_LP^S_R\cos\gamma)$. The bias voltage dependence of torque efficiency is represented in Fig.~\ref{fig:efficiency} for $T_\perp$ and $T_\|$. They both change sign with bias. These curves are in very good agreement with those obtained within the tight binding model (cf. Fig.~6(a) and (b) in Ref.~\citenum{Kalitsov2013}).

%\subsection{Torque efficiency}

\begin{figure}[t]
\includegraphics[width=\columnwidth]{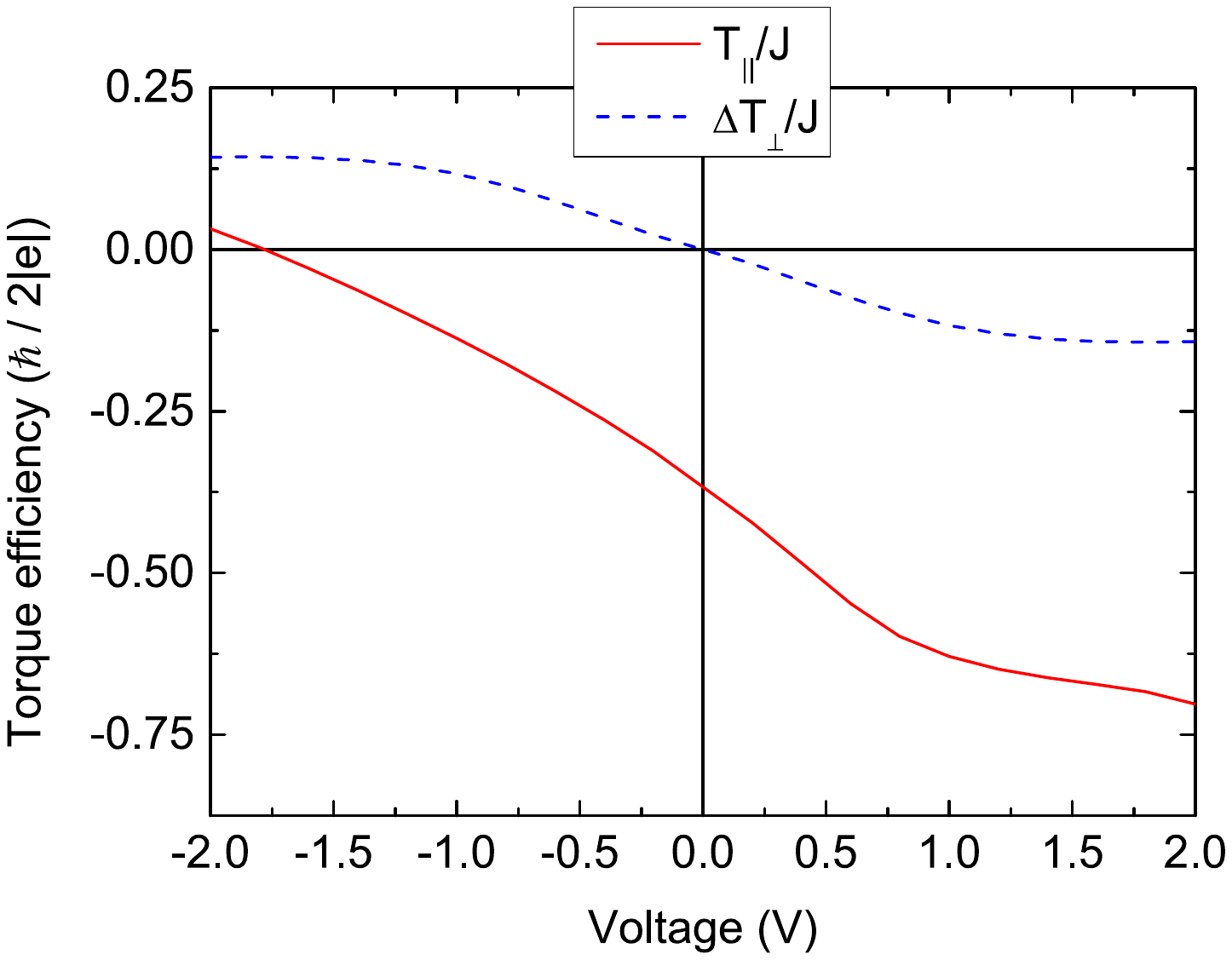}
\caption{\label{fig:Teff} Bias voltage dependence of the torque efficiency for  an MTJ. $\Delta T_\perp = T_\perp(V)-T_\perp(0)$. The parameters are: $\Delta_0=2.62$eV, $\Delta_L=\Delta_R=1.96$eV, $U_B=2$eV, $m_{\mathrm{eff}}=0.4$, $\gamma=\pi/2$, $d=7$\AA. \label{fig:efficiency}}
\end{figure}

To conclude, we derived explicit analytical formulas for spin and charge currents as well as for spin transfer torques for MTJ with non-collinear moment orientation, using only three irreducible quantities without further approximation. We showed the voltage dependence properties of STT and established conditions for deviation from the conventional sine angular dependence of both spin currents and spin torques. Furthermore, we have shown that in the large barrier approximation all tunneling transport quantities can be expressed in an extremely simplified form via Slonczewski spin polarization and the newly introduced ``effective spin-averaged interfacial transmission probabilities" and ``effective out-of-plane polarizations" at both interfaces which define the detailed angular dependence of TMR and STT. In addition, it is directly proven that for any applied voltage, the parallel component of spin current at the FM/I interface is expressed via collinear longitudinal spin current components. The developed model can be easily adapted to study thermally induced transport properties including magnetopower, thermal torques as well as other spin caloritronics phenomena by performing the energy integration exposed in the Appendix using finite temperature Fermi-Dirac distributions. We provide in Ref.~\citenum{Supplementary} the Mathematica code used in the present work.
%\subsection{Torque distribution in right FM electrode}

\appendix*
\section*{Appendix}
An expression for the current density is:
\begin{equation}
{J_e} =-\frac{|e|}{\hbar}\mathrm{Tr}(\hat J\hat I)= -\frac{|e|}{\hbar}\sum_{i,\sigma} {J_{B,i}^{\sigma \sigma}},
\end{equation}
where $i$ is ``L'' or ``R'', $\sigma$ is ``$\uparrow$''(+) or ``$\downarrow$''(-) and  the current in the barrier is given by:
\begin{equation}
J_{B,i}^{\sigma\sigma}=\sum_{\sigma'}{\Lambda_{B,i}^{\sigma(\sigma ')}}.
\end{equation}
Spin current densities and spin transfer torques in the barrier are:
\begin{align}
\begin{split}
Q_x^B &= {T_\parallel } = \frac12\mathrm{Tr}(\hat J\hat\sigma_x) = \frac12\sum_{i, \sigma} {J_{B,i}^{\sigma\, -\sigma}},\\
Q_y^B &= {T_\bot } = \frac12\mathrm{Tr}(\hat J\hat\sigma_y) = -\frac{\mathrm{i}}{2}\sum_{i, \sigma} {\sigma J_{B,i}^{\sigma\, -\sigma}},
\end{split}
\end{align}
where
\begin{equation}
J_{B,i}^{\sigma\, -\sigma}=\sum_{\sigma'}{\Xi_{B,i}^{\sigma(\sigma ')}}.
\end{equation}

The spin channel fluxes, $\Lambda$ and $\Xi$ are expressed as:
\begin{align}\label{eq:LambdaXi}
\begin{split}
\Lambda_{B,i}^{\sigma(\sigma ')} =& \frac{\mathrm{i}}{\left(2\pi\right)^3}\int\limits_{E_i^{\sigma '}}^{{E_{kT,i}}} \int\limits_0^{k_{F,i}^{\sigma '}} \kappa \,d\kappa \,dE  \\
& \times{f_i}\left( {A_i^{\sigma \sigma '}B_i^{\sigma \sigma '*} - A_i^{\sigma \sigma '*}B_i^{\sigma \sigma '}} \right) ,\\
\Xi_{B,i}^{\sigma(\sigma ')}  =& \frac{\mathrm{i}}{\left(2\pi\right)^3} \int\limits_{E_i^{\sigma '}}^{{E_{kT,i}}}\int\limits_0^{k_{F,i}^{\sigma '}}
 \kappa \,d\kappa \,dE \\
& \times{f_i}\left( A_i^{ \sigma \sigma '}B_i^{ -\sigma \sigma '*} - A_i^{ -\sigma \sigma '*}B_i^{ \sigma \sigma '} \right),
\end{split}
\end{align}
where
\begin{equation}
k_{F,i}^ \sigma  = \sqrt {\frac{{2m}}{{{\hbar ^2}}}\left( {E + {\Delta _0} + {\sigma\Delta_i} + \zeta_i \frac{eV}{2}} \right) } ,
\end{equation}
where $E$ and $\kappa$ are the electron energy and in-plane wave vector and $\Delta_{L(R)}=(U^{\downarrow}_{L(R)}-U^{\uparrow}_{L(R)})/2$ is the exchange energy in the left(right) electrode,

\begin{equation}
\zeta_i=
\begin{cases}
-1, & \mbox{if }i=\mbox{``L''}\\
1, & \mbox{if }i=\mbox{``R''}
\end{cases}
\end{equation}

\begin{equation}
E_i^\sigma  =  - {\Delta _0} - \sigma {\Delta _i} - \zeta_i eV/2,
\end{equation}

\begin{equation}
E_{kT,i} = -\zeta_i \frac{eV}{2} + \Delta_{kT},
\end{equation}
where $\Delta_{kT}$ is a smearing width for the functions $f_L$ and $f_R$ and can be taken, for example as $7kT$.

\begin{equation}
{f_i} = \frac{1}{{\exp \left( {\frac{{E + \zeta_i eV/2}}{{kT}}} \right) + 1}},
\end{equation}

\begin{align}
\begin{split}
A_L^{\sigma \sigma '} = \,& {2i\xi ^{\sigma \sigma '}}\frac{{\sqrt {k_L^{\sigma '}{q_L}} }}{\mathcal{D}}\frac{{{q_R} + ik_R^\sigma }}{{{E_n}}}\times \\
&\left[ \left( {{q_L} + ik_L^{ - \sigma '}} \right)\left( {{q_R} + ik_R^{ - \sigma }} \right)\frac{1}{{{E_n}}} \right.\\
&\biggl.- {E_n}\left( {{q_L} - ik_L^{ - \sigma '}} \right)\left( {{q_R} - ik_R^{ - \sigma }} \right) \biggr],
\end{split}
\end{align}

\begin{align}
\begin{split}
B_L^{\sigma \sigma '} = \,& {2i\xi ^{\sigma \sigma '}}\frac{{\sqrt {k_L^{\sigma '}{q_L}} }}{\mathcal{D}}\left( {{q_R} - ik_R^\sigma } \right){E_n}\times\\
&\left[ \left( {{q_L} + ik_L^{ - \sigma '}} \right)\left( {{q_R} + ik_R^{ - \sigma }} \right)\frac{1}{{{E_n}}} \right.\\
&\biggl.- {E_n}\left( {{q_L} - ik_L^{ - \sigma '}} \right)\left( {{q_R} - ik_R^{ - \sigma }} \right) \biggr],
\end{split}
\end{align}
where ${\xi ^{\sigma \sigma '}}$ is defined as:
\begin{equation}
{\xi ^{\sigma \sigma '}} =
\begin{cases}
\cos \left( {\gamma /2} \right), & \mbox{if } \sigma = \sigma '\\
\sin \left( {\gamma /2} \right), & \mbox{if  $\sigma=$``$\downarrow$'', $\sigma '=$``$\uparrow$''} \\
-\sin \left( {\gamma /2} \right), & \mbox{if  $\sigma=$``$\uparrow$'', $\sigma '=$``$\downarrow$''},
\end{cases}
\end{equation}

\begin{align}
\begin{split}
A_R^{ \downarrow  \uparrow \left( { \uparrow  \downarrow } \right)} =\,& 2{{\text{e}}^{ - ik_R^{ \uparrow \left(  \downarrow  \right)}d}}\frac{{\sqrt {k_R^{ \uparrow \left(  \downarrow  \right)}{q_R}} }}{\mathcal{D}}\times\\
&\frac{{k_L^ \uparrow  - k_L^ \downarrow }}{{{E_n}}} {q_L}\left( {{q_R} + ik_R^{ \downarrow \left(  \uparrow  \right)}} \right)\sin \gamma,
\end{split}
\end{align}

\begin{align}
\begin{split}
B_R^{ \downarrow  \uparrow \left( { \uparrow  \downarrow } \right)} =\,& 2{{\text{e}}^{ - ik_R^{ \uparrow \left(  \downarrow  \right)}d}}\frac{{\sqrt {k_R^{ \uparrow \left(  \downarrow  \right)}{q_R}} }}{\mathcal{D}} \times\\
&{E_n}{q_L}\left( {k_L^ \uparrow  - k_L^ \downarrow } \right)\left( {{q_R} - ik_R^{ \downarrow \left(  \uparrow  \right)}} \right)\sin \gamma,
\end{split}
\end{align}

\begin{align}
\begin{split}
&A_R^{ \uparrow  \uparrow \left( { \downarrow  \downarrow } \right)}{\text{ }} = 2i{{\text{e}}^{ - ik_R^{ \uparrow \left(  \downarrow  \right)}d}}\frac{{\sqrt {k_R^{ \uparrow \left(  \downarrow  \right)}{q_R}} }}{\mathcal{D}} \biggl[ \left( {{q_R} + ik_R^{ \downarrow \left(  \uparrow  \right)}} \right)\biggr.\\
&\mbox{ }\times\left( {q_L^2 + k_L^ \uparrow k_L^ \downarrow  - i{q_L}\left( {k_L^{ \uparrow \left(  \downarrow  \right)} - k_L^{ \downarrow \left(  \uparrow  \right)}} \right)\cos \gamma } \right)\frac{1}{{{E_n}}} \\
&\biggl.\mbox{ }- {E_n}\left( {{q_R} - ik_R^{ \downarrow \left(  \uparrow  \right)}} \right)\left( {{q_L} - ik_L^{ \uparrow \left(  \downarrow  \right)}} \right)\left( {{q_L} - ik_L^{ \downarrow \left(  \uparrow  \right)}} \right) \biggr],
\end{split}
\end{align}

\begin{align}
\begin{split}
&B_R^{ \uparrow  \uparrow \left( { \downarrow  \downarrow } \right)}{\text{ }} = 2i{{\text{e}}^{ - ik_R^{ \uparrow \left(  \downarrow  \right)}d}}\frac{{\sqrt {k_R^{ \uparrow \left(  \downarrow  \right)}{q_R}} }}{\mathcal{D}} \biggl[ \left( {{q_R} + ik_R^{ \downarrow \left(  \uparrow  \right)}} \right)\biggr.\\
&\mbox{ }\times\left( {{q_L} + ik_L^ \uparrow } \right)\left( {{q_L} + ik_L^ \downarrow } \right)\frac{1}{{{E_n}}} - {E_n}\left( {{q_R} - ik_R^{ \downarrow \left(  \uparrow  \right)}} \right)\\
&\mbox{ }\times\biggl.\left( {q_L^2 + k_L^ \uparrow k_L^ \downarrow  + i{q_L}\left( {k_L^{ \uparrow \left(  \downarrow  \right)} - k_L^{ \downarrow \left(  \uparrow  \right)}} \right)\cos \gamma } \right) \biggr],
\end{split}
\end{align}

\begin{align}\label{eq:En}
\begin{split}
{E_n} &= \exp \int\limits_0^d {q\left( y \right)dy} =\\
&= \exp \left[ {\frac{{2{\hbar ^2}d}}{{6{m^*}eV}}{{\left( {\frac{{{m^*}}}{m}} \right)}^3}\left( {q_L^3 - q_R^3} \right)} \right],
\end{split}
\end{align}

\begin{equation}
q(y) = \frac{1}{{m_{\mathrm{eff}}}}\sqrt {\frac{{2{m^*}}}{{{\hbar ^2}}}\left( {{U_B} - E + eV/2 - y\frac{{eV}}{d}} \right) + {\kappa ^2}},
\end{equation}

\begin{equation}\label{eq:k}
k_{i}^ \sigma  = \sqrt { k_{F,i}^{\sigma\,2} - \kappa^2} ,
\end{equation}

\begin{align}\label{eq:q}
\begin{split}
q_i =& \frac{1}{{{m_{{\text{eff}}}}}}\sqrt {\frac{{2{m^*}}}{{{\hbar ^2}}}\left( {{U_B} - E - \zeta_i eV/2} \right) + {\kappa ^2}}, 
\end{split}
\end{align}
where $U_B$ and $d$ represent respectively the barrier height and thickness with $m^*=m_{\mathrm{eff}}m_e$ being electron effective mass.
It has to note that bias voltage $V=1$Volt corresponds to $eV=1$eV.

\begin{widetext}
\begin{align}
\begin{split}
\mathcal{D} =& \left( {{q_L} + ik_L^ \uparrow } \right)\left( {{q_L} + ik_L^ \downarrow } \right)\left( {{q_R} + ik_R^ \uparrow } \right)\left( {{q_R} + ik_R^ \downarrow } \right)\frac{1}{{E_n^2}} \\
&+ \left( {{q_L} - ik_L^ \uparrow } \right)\left( {{q_L} - ik_L^ \downarrow } \right)\left( {{q_R} - ik_R^ \uparrow } \right)\left( {{q_R} - ik_R^ \downarrow } \right)E_n^2 \\
&- 2\left( {q_L^2 + k_L^ \uparrow k_L^ \downarrow } \right)\left( {q_R^2 + k_R^ \uparrow k_R^ \downarrow } \right) + 2{q_L}{q_R}\left( {k_L^ \uparrow  - k_L^ \downarrow } \right)\left( {k_R^ \uparrow  - k_R^ \downarrow } \right)\cos \gamma 
\end{split}
\end{align}
\end{widetext}

Similar expressions can be written for the right FM layer:
\begin{equation}
J_{R,i}^{\sigma\sigma}=\sum_{\sigma'}{\Lambda_{R,i}^{\sigma(\sigma ')}}.
\end{equation}
Spin current densities and spin transfer torques in the right FM layer are:
\begin{align}
\begin{split}
Q_x^R &= {T_\parallel } = \frac12\mathrm{Tr}(\hat J\hat\sigma_x) = \frac12\sum_{i, \sigma} {J_{R,i}^{\sigma\, -\sigma}},\\
Q_y^R &= {T_\bot } = \frac12\mathrm{Tr}(\hat J\hat\sigma_y) = -\frac{\mathrm{i}}{2}\sum_{i, \sigma} {\sigma J_{R,i}^{\sigma\, -\sigma}},
\end{split}
\end{align}
where
\begin{equation}
J_{R,i}^{\sigma\, -\sigma}=\sum_{\sigma'}{\Xi_{R,i}^{\sigma(\sigma ')}}.
\end{equation}

The spin channel fluxes, $\Lambda$ and $\Xi$ are expressed as:
\begin{align}\label{eq:OmegaPsi}
\begin{split}
\Lambda_{R,L}^{\sigma(\sigma ')} =& -\frac12\frac{1}{\left(2\pi\right)^3}\int\limits_{E_L^{\sigma '}}^{{E_{kT,L}}} \int\limits_0^{k_{F,L}^{\sigma '}}
\kappa \,d\kappa \,dE,\\
&\times f_L\left(k_R^\sigma+ k_R^{\sigma *}\right)\left|T^{\sigma\sigma'}\right|^2  \\
\Lambda_{R,R}^{\sigma(\sigma ')} =& -\frac12\frac{1}{\left(2\pi\right)^3}\int\limits_{E_R^{\sigma '}}^{{E_{kT,R}}} \int\limits_0^{k_{F,R}^{\sigma '}}
 \kappa \,d\kappa \,dE\\
&\times f_R(1-\left(k_R^\sigma+k_R^{\sigma *}\right)\left| R^{\sigma\sigma'}\right|^2),\\
\Xi_{R,L}^{\sigma(\sigma ')} =& -\frac12\frac{1}{\left(2\pi\right)^3} \int\limits_{E_L^{\sigma '}}^{{E_{kT,L}}} \int\limits_0^{k_{F,L}^{\sigma '}}
 \kappa \,d\kappa \,dE\\
&\times f_Lk^+_{R,\sigma}T^{\sigma\sigma'}T^{-\sigma\sigma' *}\mathrm{e}^{i\sigma k^-_{R,\sigma}y},\\
\Xi_{R,R}^{\sigma(\sigma ')} =& -\frac12\frac{1}{\left(2\pi\right)^3} \int\limits_{E_R^{\sigma '}}^{{E_{kT,R}}} \int\limits_0^{k_{F,R}^{\sigma '}}
 \kappa \,d\kappa \,dE \\
&\times f_R\Bigl[-\sigma' k^-_{R,\sigma}\tilde R^{\sigma\sigma'}\tilde R^{-\sigma\sigma'*}
\mathrm{e}^{-i\sigma\sigma' k^+_{R,\sigma}y}\Bigr.\\
\Bigl.&\mbox{ }\mbox{ }\mbox{ }+k^+_{R,\sigma}R^{\sigma\sigma'}R^{-\sigma\sigma'*}
\mathrm{e}^{i\sigma k^-_{R,\sigma}y}\Bigr],
\end{split}
\end{align}
where
\begin{equation*}
\tilde R^{\sigma\sigma'}=
\begin{cases}
\displaystyle\frac{1}{\sqrt{ k_R^\sigma}}, & \mbox{if }\sigma=\sigma'\\
R^{\sigma\sigma'}, & \mbox{if }\sigma \ne \sigma'
\end{cases}
\end{equation*}
\begin{equation*}
k^\pm_{R,\sigma}=
\begin{cases}
k^\uparrow\pm k^{\downarrow *}, & \mbox{if }\sigma=1 ("+")\\
k^{\uparrow *}\pm k^\downarrow, & \mbox{if }\sigma=-1 ("-")
\end{cases}
\end{equation*}

\begin{align}
\begin{split}
T_L^{\sigma\sigma'}=&4i\xi^{\sigma\sigma'}\frac{\sqrt{q_L q_R k_L^{\sigma'}}}{\mathcal{D}}\mathrm{e}^{-ik_R^\sigma d}\\
&\times\Biggl[\left(q_L+ik_L^{-\sigma'}\right)\left(q_R+ik_R^{-\sigma}\right)\frac{1}{E_n}\Biggr.\\
\Biggl.&-\left(q_L-ik_L^{-\sigma'}\right)\left(q_R-ik_R^{-\sigma}\right)E_n \Biggr],
\end{split}
\end{align}

\begin{align}
\begin{split}
R^{\downarrow\uparrow(\uparrow\downarrow)}=&4\mathrm{e}^{-i(k_R^\uparrow+k_R^\downarrow)d}\frac{\sqrt{k_R^{\uparrow(\downarrow)}}}{\mathcal{D}}\\
&\times q_Lq_R\left(k_L^\uparrow-k_L^\downarrow\right)\sin\gamma,
\end{split}
\end{align}

\begin{align}
\begin{split}
&R^{\uparrow\uparrow(\downarrow\downarrow)}=2i\mathrm{e}^{-2ik_R^{\uparrow(\downarrow)}d}
\frac{\sqrt{k_R^{\uparrow(\downarrow)}}}{\mathcal{D}} \Biggl[ \Biggr.\\
&\mbox{ }\Biggl. 2i\left(k_R^{\downarrow(\uparrow)}\left(q_L^2+k_L^\uparrow k_L^\downarrow\right) \mp q_Lq_R\left(k_L^\uparrow-k_L^\downarrow\right)\cos\gamma \right) \Biggr.\\
&\mbox{ }\Biggl. + \left(q_L^2-k_L^\uparrow k_L^\downarrow+iq_L\left(k_L^\uparrow + k_L^\downarrow\right)\right)\left(q_R+ik_R^{\downarrow(\uparrow)}\right)\frac{1}{E_n^2} \Biggr.\\
&\mbox{ }\Biggl. - \left(q_L^2-k_L^\uparrow k_L^\downarrow-iq_L\left(k_L^\uparrow + k_L^\downarrow\right)\right)\left(q_R-ik_R^{\downarrow(\uparrow)}\right)E_n^2 \Biggr]\\
&\mbox{ }-\frac{\textrm{e}^{-2ik_R^{\uparrow(\downarrow)}d}}{\sqrt{k_R^{\uparrow(\downarrow)}}}
\end{split}
\end{align}

\begin{acknowledgments}
We thank J.~Slonczewski, J.~Sun, O.~Mryasov, J.~Velev, X. Waintal and A.~Fert for fruitful discussions. 
This work was supported by Chair of Excellence Program of the Nanosciences Foundation in Grenoble, France and ERC Adv. Grant ``HYMAGINE'' \textnumero~246942,  Russian Fund of Basic Research grant \textnumero~13-02-01452A, and NSF-DMREF grant No. 1235396. A.M. acknowledges supports from the King Abdullah University of Science and Technology (KAUST).\end{acknowledgments}


\begin{thebibliography}{100}
\bibitem{Slonc96} J. C. Slonczewski, J. Magn. Magn. Mater. {\bf159}, L1 (1996). 

\bibitem{Berger} L. Berger, Phys. Rev. B {\bf54} 9353, (1996).

\bibitem{huai1} J.Z. Sun, J. Magn. Magn. Mater. {\bf202}, 157 (1999). 

\bibitem{huai2} Y. Huai, F. Albert, P. Nguyen, M. Pakala, and T. Valet, Appl. Phys. Lett. {\bf84}, 3118 (2004). 

\bibitem{Fuchs} G. D. Fuchs, N. C. Emley, I. N. Krivorotov, P. M. Braganca, E. M. Ryan, S. I. Kiselev, J. C. Sankey, D. C. Ralph, R. A. Buhrman, and J. A. Katine, Appl. Phys. Lett. {\bf85}, 1205
(2004). 

\bibitem{Ohno} D. Chiba, Y. Sato, T. Kita, F. Matsukura, and H. Ohno, Phys. Rev. Lett. {\bf93}, 216602 (2004).

\bibitem{chapter1} C. Baraduc, M. Chshiev, U. Ebels, {\em Introduction to spin transfer torque}, in {\em Nanomagnetism and Spintronics - Fabrication, Materials, Characterization and Applications}, Eds: F. Nasirpouri, A. Nogaret,  World Scientific Publishing, Singapore (2009).

\bibitem{chapter2} A. Manchon and S. Zhang, {\em Spin Transfer Torque: Theory}, in {\em Spin Transport and Magnetism in Electronics Systems}, Eds. E. Y. Tsymbal and I. Zutic, Taylor and Francis (2010).

\bibitem{slonc05} J. C. Slonczewski, Phys. Rev. B {\bf 71}, 024411 (2005).
 
\bibitem{slonc07} J.C. Slonczewski and J.Z. Sun, J. Magn. Magn. Mater. {\bf310}, 169-175 (2007).

\bibitem{kalitsov06} A. Kalitsov, I. Theodonis, N. Kioussis, M. Chshiev, W. H. Butler, A. Vedyayev, J. Appl. Phys. 99 (2006) 08G501.

\bibitem{theo} I. Theodonis, N. Kioussis, A. Kalitsov, M. Chshiev, and W. H. Butler, Phys.Rev. Lett. {\bf97}, 237205 (2006). 

\bibitem{Chshiev} M. Chshiev, I. Theodonis, A. Kalitsov, N. Kioussis, and W. H. Butler, IEEE Trans. Mag.  44, 2543 (2008)

\bibitem{kalitsov09} A. Kalitsov, M. Chshiev, I. Theodonis, N. Kioussis, and W. H. Butler, Phys. Rev. B {\bf79}, 174416 (2009).

\bibitem{kioussis} Y.-H. Tang, N. Kioussis, A. Kalitsov, W. H. Butler, and R. Car, Phys. Rev. Lett. {\bf103}, 057206 (2009); Phys. Rev. B {\bf81}, 054437 (2010).

\bibitem{slonc89} J.C. Slonczewski, Phys. Rev. B {\bf39}, 6995 (1989).

\bibitem{wil} M. Wilczynski, J. Barnas, and R. Swirkowicz, Phys. Rev. B {\bf77}, 054434 (2008). 

\bibitem{xiao} J. Xiao, G. E. W. Bauer, and A. Brataas, Phys. Rev. B {\bf77}, 224419 (2008).

\bibitem{manchon} A. Manchon, N. Ryzhanova, N. Strelkov, A. Vedyayev, M. Chshiev and B. Dieny, J. Phys.: Condens. Matter {\bf20}, 145208 (2008); {\em ibid} {\bf19}, 165212 (2007).

\bibitem{heiliger} C. Heiliger and M. D. Stiles, Phys. Rev. Lett. {\bf100}, 186805 (2008).

\bibitem{guo} Xingtao Jia, Ke Xia, Youqi Ke, and Hong Guo, Phys. Rev. B {\bf 84}, 014401 (2011).

\bibitem{sankey} J. C. Sankey, Y.-T. Cui, R. A. Buhrman, D. C. Ralph, J. Z. Sun, and J. C. Slonczewski, Nature Physics {\bf4}, 67
(2008)
\bibitem{kubota} H. Kubota, A. Fukushima, K. Yakushiji, T. Nagahama, S. Yuasa, K. Ando, H. Maehara, Y. Nagamine, K. Tsunekawa, D. D. Djayaprawira, N. Watanabe, and Y. Suzuki, Nature Physics {\bf4}, 37 (2008).

\bibitem{swstt} P. M. Levy and A. Fert, Phys. Rev. Lett. {\bf97}, 097205 (2006); Phys. Rev. B {\bf74}, 224446 (2006).
\bibitem{magnon} A. Manchon and S. Zhang, Phys. Rev. B {\bf79}, 174401 (2009).
\bibitem{manchon2} A. Manchon, S. Zhang, and K.-J. Lee, Phys. Rev. B {\bf82}, 174420 (2010).
\bibitem{manchon3} A. Manchon, R. Matsumoto,  H. Jaffres,  and J. Grollier, Phys. Rev. B 86, 060404(R) (2012).
\bibitem{Kalitsov2013} A. Kalitsov, W. Silvestre, M. Chshiev, and J. P. Velev, Phys. Rev. B {\bf 88}, 104430 (2013).
\bibitem{pauyac} C. Ortiz Pauyac, A. Kalitsov, A. Manchon, and M. Chshiev,  Phys. Rev. B  {\bf 90}, 235417 (2014).
\bibitem{merodio} P. Merodio, A. Kalitsov, H. Bea, V. Baltz, and M. Chshiev,  Phys. Rev. B  {\bf 90}, 224422 (2014).
\bibitem{useinov} A. Useinov, M. Chshiev and A. Manchon,  Phys. Rev. B  {\bf 91}, 064412 (2015).
\bibitem{deac} A. M. Deac, A. Fukushima, H. Kubota, H. Maehara, Y. Suzuki, S. Yuasa, Y. Nagamine, K. Tsunekawa, D. D. Djayaprawira, N. Watanabe, Nature Physics {\bf4}, 803 (2008).
\bibitem{petit} S. Petit, C. Baraduc, C. Thirion, U. Ebels, Y. Liu, M. Li, P. Wang, and B. Dieny, Phys. Rev. Lett. {\bf98}, 077203 (2007).
\bibitem{li} Z. Li, S. Zhang, Z. Diao, Y. Ding, X. Tang, D.M. Apalkov, Z. Yang, K. Kawabata, and Y. Huai, Phys. Rev. Lett. {\bf100}, 246602 (2008).
\bibitem{sun} J. Z. Sun, M. C. Gaidis, G. Hu, E. J. O'Sullivan, S. L. Brown, J. J. Nowak, P. L. Trouilloud, and D. C. Worledge, J. Appl. Phys. {\bf105}, 07D109 (2009); T. Min, J. Z. Sun, R. Beach, D. Tang, and P. Wang, J. Appl. Phys. {\bf105}, 07D126 (2009).
\bibitem{oh} S.-C. Oh, S.-Y. Park, A. Manchon, M. Chshiev, J.-H. Han, H.-W. Lee, J.-E. Lee, K.-T. Nam, Y. Jo, Y.-C. Kong, B. Dieny,and K.-J. Lee, Nature Physics {\bf 5}, 898 (2009).
\bibitem{chanthbouala} A. Chanthbouala, R. Matsumoto, J. Grollier, V. Cros, A. Anane, A. Fert, A. V. Khvalkovskiy, K.A. Zvezdin, K. Nishimura, Y. Nagamine, H. Maehara, K. Tsunekawa, A. Fukushima and S. Yuasa, Nature Physics {\bf 7}, 626 (2011).
\bibitem{Supplementary} See Supplemental Material at [URL will be inserted by publisher] with the code used in the present work.
\bibitem{stearns} M. B. Stearns, J. Magn. Magn. Mater. {\bf5}, 167 (1977).
\bibitem{ZhangButler} X.-G. Zhang and W. H. Butler, J. Phys.: Condens. Matter {\bf 15} (2003) R1603
\bibitem{BelaschTsymbal} K. D. Belashchenko, E. Y. Tsymbal, M. van Schilfgaarde, D. A. Stewart, I. I. Oleinik and S. S. Jaswal, Phys. Rev. B 69, 174408 (2004)
\bibitem{IEC_exp} J. Faure-Vincent, C. Tiusan, C. Bellouard, E. Popova, M. Hehn, F. Montaigne, and A. Schuhl, Phys. Rev. Lett. 89, 107206 (2002).
\bibitem{IEC_the} M. Y. Zhuravlev, J. Velev, A. V. Vedyayev, and E. Y. Tsymbal, J. Magn. Magn. Mater. 300, e277 (2006).
\bibitem{Yang} H. X. Yang et al., Appl. Phys. Lett. 96, 262509 (2010).
\bibitem{Stiles} M.~D.~Stiles and A.~Zangwill, Phys. Rev. B 66, 014407 (2002)
\bibitem{Slon2002} J.~C.~Slonczewski, J. Magn. Magn. Mater. 247, 324 (2002)
\bibitem{Waintal2010} C.~Petitjean, D.~Luc, and X.~Waintal, Phys. Rev. Lett. 109, 117204 (2010)
\bibitem{note} Similar length scales has been introduced for metallic spin valves in Ref.~\citenum{Waintal2010}

\end{thebibliography}
\end{document}